\documentclass[aps,prd,twocolumn,groupedaddress,showpacs,showkeys]{revtex4-1}

\usepackage{graphicx}

\begin{document}



\title{ARCHIMEDEAN-TYPE FORCE IN A COSMIC DARK FLUID: \\
I. EXACT SOLUTIONS FOR THE LATE-TIME ACCELERATED EXPANSION}


\author{Alexander B. Balakin\footnote{e-mail: Alexander.Balakin@ksu.ru} and Vladimir V.
Bochkarev\footnote{e-mail: Vladimir.Bochkarev@ksu.ru}}
\affiliation{Kazan Federal University, Kremlevskaya str.,
18, 420008, Kazan,  Russia}

\date{\today}

\begin{abstract}
We establish a new self-consistent model in order to explain from
a unified viewpoint two key features of the cosmological evolution:
the inflation in the early Universe and the late-time accelerated
expansion. The key element of this new model is the
Archimedean-type coupling of the dark matter with dark energy,
which form the so-called cosmic dark fluid. We suppose that
dark matter particles immersed into the dark energy reservoir are
affected by the force proportional to the four-gradient
of the dark energy pressure. The Archimedean-type coupling  is
shown to play a role of effective energy-momentum redistributor
between the dark matter and the dark energy components of the dark fluid,
thus providing the Universe evolution to be a quasiperiodic
and/or multistage process. In the first part of the work we
discuss a theoretical base and new exact solutions of the model
master equations. Special attention is focused on the exact
solutions, for which the scale factor is presented by the
anti-Gaussian function: these solutions describe the late-time
acceleration and are characterized by a nonsingular behavior in
the early Universe. The second part contains qualitative and
numerical analysis of the master equations; we focus there on the
solutions describing a multi-inflationary Universe.
\end{abstract}


\maketitle

\section{Introduction}

The concepts of dark energy (DE) and dark matter (DM) (see, e.g.,
\cite{DE1,DE2,DE3} and  \cite{DM1,DM2,DM3} for review and
references) are the basic elements of modern cosmology and
astrophysics. These elements were introduced into the scientific lexicon in
two different ways: dark energy is considered to be a reason
for the late-time accelerated expansion of the Universe
\cite{A1,A2}, while dark matter is usually associated with the
explanation of the flat velocity curves of the spiral galaxies
rotation \cite{Sp1,Sp2}. Nevertheless, there exists a tendency to
consider dark energy and dark matter as two manifestations of one
unified dark fluid (see, e.g., \cite{DF1,Oreview,DF2,DF3,DF4,DF5}). The
models of interaction between two constituents of the dark fluid,
as well as the models of interactions  of dark energy and/or dark
matter   with the standard (baryon) matter, are the subjects of wide
discussion. The total contribution of dark energy and dark matter
into the energy balance of the Universe is estimated to be about
$95 \%$. Thus, the coupling between these two constituents of the
dark fluid seems to be the most important element in the list of
cosmic medium interactions, and the dark fluid can be considered
as a thermodynamic reservoir for baryon matter.

The most developed model of the coupling between DM and DE
constituents of the dark fluid is based on the two-fluid
representation of the cosmic
medium (see, e.g., \cite{M1,M2,M3,M4,M5,M6}). In this approach the interaction
terms $\pm Q$ appear in the
right-hand sides of separate balance equations for the DE and DM
with opposite signs and disappear in a sum, when one deals with
the total balance equation. The modeling of the coupling term $Q$
has in most cases a phenomenological character and is based on the
ansatz that $Q$ is a function (e.g., linear or power-law) of the
energy densities of the DE and DM, of the Hubble function, $H$,
etc. We formulate the theory of interaction between DE and DM
using the relativistic hydrodynamics for dark energy and
relativistic kinetics for dark matter. We suggest that the DE
acts on the DM particles by means of some effective force, and the
corresponding model force four-vector is introduced into the
kinetic equation. The backreaction of the DM on the DE is
described using a force-type term in the hydrodynamic equations. The
total system (DM plus DE) is considered to be conservative.
The concept of the Archimedean-type force can be naturally
generalized for the description of the DE action on the baryon matter; however,
now we restrict ourselves by the model of the DE and DM interaction.

We discussed the structure and properties of various effective forces, which appeared in the cosmological contexts, in
the papers \cite{BZ1,BZ2,BZ3,BZ4,BZ5,BZ6}, the relativistic generalizations of the Stokes force, Langevin force,
antifriction and tidal forces being investigated in detail.
Concerning the force acting on the DM particles from the DE we would like to introduce the so-called
Archimedean-type force. This choice can be motivated as follows.
{\it First} of all, this force is a relativistic generalization of the classical Archimedean force, proportional to the
three-gradient of the Pascal pressure; thus, the model under discussion is based on the well-known and well examined scheme of interaction.
{\it Second}, the DE pressure is assumed to be of the same order as the
DE energy density (73\% of the Universe energy density), thus, the Archimedean effect on the dark matter could be
significant. {\it Third}, assuming that the DE pressure can be negative,
we obtain that such a force can be attractive in contrast to expulsive
classical Archimedean force. Admitting that the DE pressure changes the sign in the course of the Universe evolution, one can describe a multistage (or
even quasiperiodic) character of the cosmological expansion, for which epochs of deceleration are changed by epochs of acceleration and vice versa. (The interest
in models of this type was renewed by the paper \cite{Penrose}).

We show that, in principle, the Archimedean-type force can effectively redistribute $95\%$ of the
Universe's energy between the DE and DM constituents, thus guiding the time evolution
of the cosmic medium as a whole. We divided the work into two parts: the first one contains pure analytical results and exact solutions of the model;
in the second part we focus on the numerical and qualitative analysis of the model.

The first part of the work is organized as follows. In Sec.II
we derive the master equations of the model with an
Archimedean-type force. In particular, in Sec. II.B based on the kinetic
approach we introduce the Archimedean-type force, obtain basic
integrals of motion, construct the distribution functions and
calculate their macroscopic moments as functions of the DE
pressure. In Sec. II.C we formulate the balance equation for
the dark fluid. In Sec. II.D we discuss the extended
(inhomogeneous) equation of state for the dark energy and
introduce the key equation for the DE pressure evolution. Sec. III
contains discussions about two classes of exact solutions. In
Sec. III.A we consider a special (constant) exact solution
for the case when the guiding parameter of the model $\sigma$
is not equal to its critical value, i.e. $\sigma \neq -1$, and
we analyze the problem of asymptotic stability of this
solution. In Sec. III.B we focus on the special case $\sigma
= -1$ and obtain exact solutions of the  anti-Gaussian type for
the following submodels: (i) the massless DM, (ii) the cold dark
matter, and (iii) the submodel with the DE domination. In Sec. IV
we discuss obtained analytical results.

\section{Master equations}

We consider a cosmic medium, which consists of two interacting
components. One of them, the dark matter, can be
described in the framework of general relativistic kinetic theory
\cite{kin1,kin2}. The second component, dark energy, is considered
as a perfect fluid with inhomogeneous equation of state. These two components interact gravitationally,
i.e., both of them contribute the energy and momentum terms into the
total stress-energy tensor of the system as a whole, indicated as the dark fluid.
In addition, we assume that DE and DM interact by means of force of the Archimedean type. Mathematically,
this model can be described as follows.

\subsection{Equations for gravity field}

We consider the spatially homogeneous  Friedmann-Lema$\hat{i}$tre-Robertson-Walker (FLRW) cosmological model
with the metric
\begin{equation} ds^2 = dt^2 -a^2(t)[(dx^1)^2 + (dx^2)^2
+(dx^3)^2] \,. \label{metric}
\end{equation}
Here and below $c=1$. The Einstein equations
\begin{equation}
R_{ik} - \frac12 R g_{ik} = \kappa T_{ik}^{({\rm total})} +
\Lambda g_{ik}\,, \label{Ein}
\end{equation}
can be reduced to the well-known system
\begin{equation}
\frac{\ddot{a}}{a} = - \frac{4\pi G}{3}[{\cal E} + 3 {\cal P}] +
\frac{\Lambda}{3} \,, \qquad \left(\frac{\dot{a}}{a} \right)^2 =
\frac{8\pi G}{3}{\cal E} + \frac{\Lambda}{3} \,,
\label{EinREDU}
\end{equation}
where the dot denotes the derivative with respect to time, ${\cal
E}(t)$ is the total energy of the system as a whole, and ${\cal
P}$ is the corresponding pressure. Since the cosmological constant
$\Lambda$ is frequently interpreted in terms of vacuum energy and
can be considered as a dark energy candidate, we assume the
following decompositions of the total energy and pressure
\begin{equation}
{\cal E} + \frac{\Lambda}{8\pi G} = \rho + E \,, \quad {\cal P} -
\frac{\Lambda}{8\pi G} = \Pi + P \,. \label{Decomp1}
\end{equation}
In such decomposition the functions  $\rho(t)$ and $\Pi(t)$
describe the dark energy, while the functions $E(t)$ and $P(t)$
describe the dark matter; thus, the term $\frac{\Lambda}{8\pi G}$ is
considered to be incorporated into the DE energy density $\rho$ and DE
pressure $\Pi$. Now the master equations for gravity field are
\begin{equation}
\frac{\ddot{a}}{a} = - \frac{4\pi G}{3}[ (\rho + E) + 3 (\Pi + P)]
\,,  \label{EinREDU10}
\end{equation}
\begin{equation}
H^2 = \frac{8\pi G}{3}(\rho + E) \,, \label{EinREDU1}
\end{equation}
where, as usual, $H(t) \equiv \frac{\dot{a}}{a}$ is the Hubble
function.

\subsection{Kinetic equation for relativistic DM particles}

Let us consider the general relativistic kinetic equations
\begin{equation} \frac{p^i}{m_{({\rm a})}} \left(
\frac{\partial}{\partial x^i} - \Gamma^k_{il} p^l
\frac{\partial}{\partial p^k}\right) f_{({\rm a})} + \frac{\partial}{\partial
p^i} \left[{\cal F}_{({\rm a})}^i f_{({\rm a})} \right] = 0  \label{KinEq}
\end{equation}
for the distribution functions $f_{({\rm a})}$, describing the
evolution of the DM particles with the masses $m_{({\rm a})}$. It is worth mentioning
that the DM can consist of a few sorts of particles (massless and massive), thus, we use the index ${({\rm
a})}$ to distinguish them. The kinetic equations (\ref{KinEq}) are of
the collisionless type, i.e., we neglect direct interparticle
collisions, but consider the force four-vector ${\cal F}^i_{({\rm
a})}$ to guide the DM particle dynamics. Two subsets of
characteristic equations
\begin{equation}
\frac{dp^k}{ds} + \Gamma^k_{jl} p^j \frac{dx^l}{ds} = {\cal F}^k_{({\rm a})}
\,, \quad \frac{dx^i}{ds} = \frac{p^i}{m_{({\rm a})}} \,,
\label{Character}
\end{equation}
show that the DM particle motion is not geodesic because of the force ${\cal
F}^i_{({\rm a})}$, which appears as a result of the DE action. The
third subset of characteristic equations
\begin{equation}
\frac{df_{({\rm a})}}{ds} = - f_{({\rm a})} \left(\frac{\partial {\cal F}^k_{({\rm a})}}{\partial p^k}
\right)  \label{Character3}
\end{equation}
demonstrates the important role of the divergence $\frac{\partial
{\cal F}^k_{({\rm a})}}{\partial p^k}$ in the evolution of the distribution
functions.

\subsubsection{ Archimedean-type force}

In classical physics the term ``Archimedean force" appears when one
deals with a body immersed into a nonhomogeneous liquid. This
three-dimensional force $\vec{F}_{({\rm Arch})}$ is proportional to the spatial
gradient of the Pascal pressure $P_{({\rm Pascal})}$
\begin{equation} \vec{F}_{({\rm Arch})} = - V_0 \vec{\nabla} P_{({\rm Pascal})}
\,. \label{force0}
\end{equation}
Clearly, in the Archimedean historical experiment the gradient of the
pressure was equal to the product of the water mass density $\rho_{({\rm water})}$
and the free-fall acceleration $g$ in Syracuse. The coefficient $V_0$ relates to a body volume
and can be represented as a quotient $\frac{M_{({\rm body})}}{\rho_{({\rm body})}}$,
where $M_{({\rm body})}$ is a mass of the body
and $\rho_{({\rm body})}$ is its mass density.
The minus sign corresponds to the fact that the Archimedean force is the
buoyancy (expulsive) one.

This classical force can be generalized for the
case of relativistic DM particle immersed into the DE reservoir.  The
force four-vector
\begin{equation} {\cal F}^i_{({\rm a})} = m_{({\rm a})} {\cal
V}_{({\rm a})} \left[g^{ik} - \frac{p^i p^k}{p^l p_l} \right] \nabla_k \Pi
 \label{force}
\end{equation}
can be considered as a generalization. First of all, it
is a force proportional again to the gradient of pressure $\Pi$,
but now we deal with a four-gradient instead of gradient
three-vector in classical physics, and $\Pi$ is the DE pressure instead of the Pascal one.
Second, the projector in the
square brackets guarantees that the force four-vector is
orthogonal to the particle momentum four-vector, i.e., ${\cal F}^i_{({\rm a})}
p_i {=}0$. The last property provides the existence of the first integral of motion
$p^k p_k {=} m^2_{({\rm a})} {=} const$, which guarantees the particle mass conservation.
New constants ${\cal V}_{({\rm a})}$ have the
dimensionality of inverse energy density (remember that $c{=}1$) and
describe some new effective constants of interaction. The sign in front of these constants is
positive, and this is in agreement with the definition
(\ref{force0}) and the accepted signature ($+---$) of the metric
(\ref{metric}). This force does not belong to the class of
gyroscopic forces, for which $\frac{\partial {\cal F}^k_{({\rm a})}}{\partial
p^k}{=} 0$. Indeed, simple calculations give
\begin{equation}
\frac{\partial {\cal F}^k_{({\rm a})}}{\partial p^k} = - 3 m_{({\rm a})} {\cal V}_{({\rm a})} \frac{
p^k \nabla_k \Pi}{p^l p_l} = - 3{\cal V}_{({\rm a})} \frac{ p^0
\dot{\Pi}}{m_{({\rm a})}}  \neq 0 \,, \label{diva}
\end{equation}
where $p^0$ is the DM particle energy
\begin{equation}
p^0=p_0 = \sqrt{m^2_{({\rm a})}-p^{\alpha} p_{\alpha}} \,. \label{1integral}
\end{equation}
(Greek indices run from 1 to 3).
Thus, the Archimedean-type force (\ref{force}) is divergence-free, if and only if the DE pressure $\Pi$ is constant.

\subsubsection{Integrals of motion}

Using the force four-vector given by (\ref{force}) one can extract
the following self-closed subsystem from the characteristic
equations (\ref{Character}):
\begin{equation}
\frac{dp_{\alpha}}{dt} = - {\cal V}_{({\rm a})} p_{\alpha}
 \dot{\Pi}\,, \quad \frac{dt}{ds} = \frac{p^0}{m_{({\rm a})}} \,. \label{Char1}
\end{equation}
These equations yield immediately three first integrals (for the
covariant components of the particle momentum)
\begin{equation}
p_{\alpha}(t) {=} C_{\alpha} \exp\left\{{\cal
V}_{({\rm a})} \left[\Pi(t_0){-}\Pi(t) \right]\right\} , \
p_{\alpha}(t_0) {=} C_{\alpha} \,. \label{impuls}
\end{equation}
Thus, any deviation of the DE pressure $\Pi(t)$ from the initial
value $\Pi(t_0)$ generates particle acceleration/deceleration due
to the Archimedean-type interaction.
The energy of the DM particle of the sort $({\rm a})$
\begin{equation}
p^0 {=} m_{({\rm a})} \sqrt{ 1 {+} q^2 \left[\frac{a(t_0)}{a(t)}\right]^2
e^{2 {\cal V}_{({\rm a})} \left[\Pi(t_0){-}\Pi(t)
\right]} }  \label{energy}
\end{equation}
deviates from its initial value
\begin{equation}
p^0(t_0) =  m_{({\rm a})} \sqrt{1+q^2} \,, \label{q002}
\end{equation}
where
\begin{equation}
q^2 \equiv \frac{1}{m^2_{({\rm a})} a^2(t_0)} \left[ C_1^2 + C_2^2 + C_3^2
\right]\,, \label{q2}
\end{equation}
due to the following effects. The first effect is the standard energy
decreasing due to the Universe's expansion; it is described by the
term $a^2(t_0) / a^2(t)$ and tends to make the matter effectively
nonrelativistic at $t \to \infty$. A new effect caused by the
nonstationarity of the dark energy is described in (\ref{energy})
by the exponential term. Let us emphasize, that this effect
decreases the particle energy, if the term ${\cal V}_{({\rm a})}
\left[\Pi(t){-}\Pi(t_0)\right]$ is positive, and increases it in
the opposite case. Let us consider only one simple example, when
${\cal V}_{({\rm a})}$ is positive, the initial value $\Pi(t_0)$
is vanishing: one can see, that the DM particle energy grows due
to the exponential term, when the DE pressure is negative. In this
sense, the Archimedean-type force with negative DE pressure
counteracts the effective cooling of the dark matter in the course
of expansion. In other words, even if a particle has small (but
not equal identically to zero) initial kinetic energy, it can
become ultrarelativistic due to the Archimedean-type interaction
with the dark energy.

\subsubsection{Macroscopic moments of the distribution function}

Solving Eq. (\ref{Character3}) taking into account (\ref{diva}) and
(\ref{impuls}) we obtain
\begin{equation}
f_{({\rm a})}(t, p_{\alpha}) =
f_{({\rm a})}^0 (q^2) \exp\left\{ 3 {\cal V}_{({\rm a})} \left[\Pi(t)-\Pi(t_0)
\right]\right\} \,, \label{f}
\end{equation}
where $f_{({\rm a})}^0(q^2)$ is an initial distribution function of the DM particles of the sort $(a)$.
Using (\ref{f}) and (\ref{impuls}) one can reduce the DM particle
stress-energy tensor (the macroscopic moment of the second order)
\begin{equation}
T_{ik}^{({\rm DM})}(t) = \sum_{({\rm a})} \int \frac{dp_1 dp_2 dp_3}{\sqrt{-g} \
p_0} f_{({\rm a})}(t, p_{\alpha}) \  p_i p_k   \label{TEI}
\end{equation}
to the pair of basic integrals for the energy $E$ and pressure
$P$, respectively:
\begin{equation}
E(x) {=} \sum_{({\rm a})}\frac{4\pi m^4_{({\rm a})}}{x^3} \int_0^{\infty} q^2 dq f_{({\rm a})}^0(q^2)
\sqrt{1{+}q^2 F_{({\rm a})}(x)} \,, \label{E(x)}
\end{equation}
\begin{equation}
P(x) {=}  \sum_{({\rm a})}\frac{4\pi m_{({\rm a})}^4 F_{({\rm a})}(x)}{3x^3} \int_0^{\infty} \frac{q^4 dq
f_{({\rm a})}^0(q^2)}{\sqrt{1{+}q^2 F_{({\rm a})}(x)}} \,. \label{P(x)}
\end{equation}
Here for the sake of convenience we have introduced a new dimensionless
variable $x$ and an auxiliary function $F_{({\rm a})}(x)$ defined as
follows
\begin{equation}
x = \frac{a(t)}{a(t_0)} \,,  \ F_{({\rm a})}(x) {=} \frac{1}{x^2}
\exp\left\{2{\cal V}_{({\rm a})} \left[\Pi(1){-}\Pi(x)
\right]\right\} . \label{Fdefin}
\end{equation}
The initial moment $t{=}t_0$ corresponds to the value $x{=}1$
($F_{({\rm a})}(1){=}1$). The DE pressure can be also rewritten in these terms
as
\begin{equation}
\Pi(x) = \Pi(1) - \frac{1}{2 {\cal V}_{({\rm a})}} \log{[x^2 F_{({\rm a})}(x)]}\,.
\label{FPI}
\end{equation}
The most conventional model of matter deals with the relativistic
Maxwell - Boltzmann functions \cite{kin2}
\begin{equation}
f_{({\rm a})}^0(q^2) = \frac{N_{({\rm a})} \lambda_{({\rm a})}}{4\pi m_{({\rm a})}^3 K_2(\lambda_{({\rm a})})}
e^{-\lambda_{({\rm a})} \sqrt{1+q^2}} \,, \label{fequi}
\end{equation}
where $N_{({\rm a})}$ is the particle number density,
$\lambda_{({\rm a})} {=} \frac{m_{({\rm a})}}{k_{({\rm B})}T_{({\rm a})}}$,
and $K_2(\lambda_{({\rm a})})$ is the modified
Bessel function \cite{kin2}, defined as
\begin{equation}
K_{\nu}(\lambda_{({\rm a})}) \equiv \int_0^{\infty} dz \cosh{\nu z} \cdot
\exp{[-\lambda_{({\rm a})} \cosh z]}   \,, \label{McD}
\end{equation}
$k_{({\rm B})}$ is the Boltzmann constant.
We assume that there is no thermodynamic equilibrium between
different sorts of DM particles, thus, generally, the temperatures $T_{({\rm
a})}$ do not coincide and are marked by the index of the sort. In
this model the DM energy and pressure have, respectively, the form
\begin{equation}
E(x) {=} \sum_{({\rm a})}\frac{E_{({\rm a})}}{x^3} \int_0^{\infty} q^2 dq \sqrt{1{+}q^2 F_{({\rm a})}(x)} \
e^{{-}\lambda_{({\rm a})} \sqrt{1{+}q^2}}\,, \label{e(x))}
\end{equation}
\begin{equation}
P(x) {=} \sum_{({\rm a})}\frac{E_{({\rm a})}}{3x^3} \int_0^{\infty} \frac{F_{({\rm a})}(x) q^4
dq}{\sqrt{1{+}q^2 F_{({\rm a})}(x)}} \ e^{{-}\lambda_{({\rm a})} \sqrt{1{+}q^2}}\,,  \label{p(x)}
\end{equation}
where the constant $E_{({\rm a})}$ is given by the formula
\begin{equation}
E_{({\rm a})}
\equiv \frac{N_{({\rm a})} m_{({\rm a})} \lambda_{({\rm a})}}{K_2(\lambda_{({\rm a})})}\,. \label{Ea}
\end{equation}
The formulas for the Fermi-Dirac and Bose-Einstein functions
can be obtained analogously, and we do not write them here.

\subsubsection{Massless particles $m_{({\rm a})}=0$}

The dark matter can consist of massless and massive particles.
When we deal with massless particles, i.e., $m_{({\rm a})}{=}0$, e.g., for the index $({\rm a}) {=} (0)$,
we have to replace formally the quantity $p^i/m_{({\rm a})}$ by $k^i$ with $k^i k_i {=}0$.
Respectively, the formula (\ref{Ea}) has to be modified as follows:
\begin{equation}
E_{(0)} \to  \frac{4\pi \tilde{\nu} [k_{({\rm B})}T_{(0)}]^4}{h^3}
\int_0^{\infty} \frac{q^3 dq}{e^q \pm 1} \,. \label{m00}
\end{equation}
Here $h$ is the Planck constant, $T_{(0)}$ is an initial
temperature of the massless particles, and $\tilde{\nu}$ is a
degeneracy factor. Clearly, this massless constituent of the DM is
described by the ultrarelativistic equation of state
\begin{equation}
E_{(0)}(x) = 3 P_{(0)}(x) = E_{(0)} \ \frac{\sqrt{F_{(0)}(x)}}{x^3} \,.
\label{m000}
\end{equation}
When the Archimedean-type force is absent, one obtains that $F_{(0)}(x){=} x^{{-}2}$ and
$E_{(0)}(x) \propto x^{{-}4}$, as it should be.

\subsection{Balance equations}

\subsubsection{Energy balance for interacting DE and DM}

We consider the model in which dark matter and dark energy form a coupled conserved system.
The total stress-energy tensor of DM and DE is divergence-free
\begin{equation}
T^{ik}_{({\rm total})} = T^{ik}_{({\rm DM})} + T^{ik}_{({\rm DE})} \,, \quad
\nabla_k T^{ik}_{({\rm total})} = 0 \,.
\label{balance000}
\end{equation}
We suppose that both cosmic substrates, DM and DE, have the same
macroscopic velocity four-vector, $U^i {=} \delta^i_0$, thus, due to
the FLRW space-time symmetry only one equation among
(\ref{balance000}) is nontrivial:
\begin{equation}
\dot{\rho} + \dot{E} + 3H (\rho + E + \Pi + P) =0 \,.
\label{balance}
\end{equation}
It is the direct differential consequence of Eqs. (\ref{EinREDU10}) and (\ref{EinREDU1}).

\subsubsection{Balance equations for dark matter}

Balance equations for DM  can be obtained  by integration of the
kinetic equation (\ref{KinEq}) (see \cite{kin1,kin2} for details).
The first-order macroscopic moment $N^k$ is defined as
\begin{equation}
N^k \equiv \sum_{({\rm a})} N^k_{({\rm a})} =
\sum_{({\rm a})} \int dP  f_{({\rm a})} p^k
\label{balanceDM1}
\end{equation}
and describes the total DM particle number four-vector. Using the kinetic equation
(\ref{KinEq}) the four-divergence of this vector can be easily calculated:
$$
\nabla_k N^k  = \nabla_k \sum_{({\rm a})} \int dP p^k f_{({\rm a})} =
$$
\begin{equation}
= {-} \sum_{({\rm a})} m_{({\rm a})}\int dP \frac{\partial}{\partial p^k} \left[f_{({\rm a})} {\cal F}^k_{({\rm a})}\right]
= 0 \,.
\label{balanceDM2}
\end{equation}
This means that the DM particle number is conserved. Taking into account the symmetry of the model one can write
\begin{equation}
N^k = \delta^k_0 N(t_0) \left[\frac{a(t_0)}{a(t)} \right]^3 \,.
\label{balanceDM3}
\end{equation}
The second-order macroscopic moment, the DM stress-energy tensor
$T^{ik}_{({\rm DM})}$, satisfies the following equation
$$
\nabla_k T^{ik}_{({\rm DM})} \equiv \nabla_k \sum_{({\rm a})} T^{ik}_{(a)} {=}
\sum_{({\rm a})} m_{({\rm a})} \int dP f_{({\rm a})} {\cal F}_{({\rm a})}^{i}
{=}
$$
$$
{=} \nabla_k \Pi \sum_{({\rm a})} m^2_{({\rm a})} {\cal V}_{({\rm a})} \int dP f_{({\rm
a})} \left[g^{ik} {-} \frac{p^i p^k}{p^l p_l} \right] {=}
$$
\begin{equation}
=
\delta^i_0 \ 3 \dot{\Pi} \sum_{({\rm a})} {\cal V}_{({\rm a})}   P_{({\rm a})}
\,.
\label{balance011}
\end{equation}
Only one equation among (\ref{balance011}) is nontrivial:
\begin{equation}
\dot{E} + 3H (E + P) = - {\cal Q} \,.
\label{balance1alt1}
\end{equation}
The source term in the right-hand side of this equation has the
form
\begin{equation}
{\cal Q} \equiv 3 \ \dot{\Pi} \sum_{({\rm a})} {\cal V}_{({\rm a})}  P_{({\rm a})} \,.
\label{balance1alt2}
\end{equation}
The source ${\cal Q}$ in the DM energy balance equation vanishes, when the DE pressure $\Pi$ is constant and the
Archimedean-type force disappears.

The scalar of the DM entropy production
\begin{equation}
\sigma_{({\rm DM})} {=} \nabla_k S^k {=} {-} k_{({\rm B})} \nabla_k \sum_{({\rm a})}
\int dP \ p^k f_{({\rm a})} \left[\log{h^3 f_{({\rm a})}} {-} 1 \right]
\label{balance33}
\end{equation}
is also proportional to the four-gradient of the DE pressure
$$
\sigma_{({\rm DM})} {=}  k_{({\rm B})} \sum_{({\rm a})} m_{({\rm a})}
\int dP f_{({\rm a})}  \frac{\partial {\cal F}^k_{({\rm a})}}{\partial
p^k} {=}
$$
$$
{=}{-} 3k_{({\rm B})} \nabla_k \Pi \sum_{({\rm a})} {\cal V}_{({\rm a})} N^k_{({\rm a})} {=}
$$
\begin{equation}
= {-} 3k_{({\rm B})} \ \dot{\Pi} \left[\frac{a(t_0)}{a(t)} \right]^3
\sum_{({\rm a})} {\cal V}_{({\rm a})} N_{({\rm a})}(t_0)
. \label{balance323}
\end{equation}
When the coefficient $\Re \equiv \sum_{({\rm a})} {\cal V}_{({\rm a})} N_{({\rm a})}(t_0)$ is positive and
$\dot{\Pi}>0$, the entropy production scalar $\sigma_{({\rm DM})}$ is negative, i.e., the DE provides organization processes
in the DM system by the Archimedean-type force; when $\dot{\Pi}<0$ the
Archimedean-type force produces a chaotization in the DM system.

\subsubsection{Balance equations for dark energy}

The combination of Eqs. (\ref{balance}) and (\ref{balance1alt1}) gives the
following balance equation for the DE state functions $\rho$ and $\Pi$:
\begin{equation}
\dot{\rho} + 3H (\rho + \Pi) =  {\cal Q}
\,. \label{balance22}
\end{equation}
The functions $P_{({\rm a})}(t)$ depend on $\Pi(t)$ according to
Eqs. (\ref{p(x)}) and (\ref{Fdefin}) for the massive
particles, and according to Eq. (\ref{m000}) for the massless ones. One can
emphasize that the macroscopic balance equations
(\ref{balance1alt1}), (\ref{balance1alt2}) and (\ref{balance22})
look like the balance equations in the well-known two-fluid models
\cite{M1,M2,M3,M4,M5}; the difference is that now the source term
${\cal Q}$ (\ref{balance1alt2}) is not modeled
phenomenologically, but is directly calculated on the basis of kinetic
approach.

\subsection{Dark energy dynamics}

\subsubsection{Dark energy equation of state accounting for
retardation of response}

To describe the dark energy fluid we use the linear equation of state
\begin{equation}
\rho(t) = \rho_0 + \sigma \Pi + \frac{\xi}{H(t)} \dot{\Pi} \,.
\label{simplest0}
\end{equation}
It belongs to the class of the so-called inhomogeneous equations
of state, which is intensely discussed in the literature (see,
e.g., \cite{EOS1,EOS2,EOS3,EOS4,EOS5,wt1}).
When $\sigma {=}0$ and $\xi{=}0$, Eq. (\ref{simplest0}) introduces the
model in which the dark energy relates to the cosmological constant
$\Lambda$ [see, e.g., (\ref{Decomp1})].
Alternatively, $\rho_0$ can be introduced by analogy with the so-called bag
constant appearing in the theory of quark-gluon plasma \cite{bag}.
When $\rho_0{=}0$ and $\xi{=}0$, Eq. (\ref{simplest0}) gives the well-known linear relation $\Pi {=} w \rho$
with $w \equiv \frac{1}{\sigma}$.
Since the proportionality coefficient $w$ depends on the choice of the
epoch in the Universe's evolution, many authors consider it as a function of cosmological time, i.e.,
$w=w(t)$, thus introducing the nonstationary equation of state (see, e.g., \cite{EOS1,EOS2,EOS3,EOS4,EOS5,wt1}).
We follow another version of nonstationary equation of state, for which $w$ and $\sigma$ remain constant, but
the retardation of response is taken into account by inserting the term containing the first derivative of the
pressure $\dot{\Pi}$. An equivalent scheme is widely used in the extended thermodynamics and rheology [see,
e.g., \cite{REO1}], in which the extended constitutive equation for the
thermodynamically coupled variables ${\bf X}$ and ${\bf Y}$ has the form
\begin{equation}
\tau \dot{{\bf X}} + {\bf X} = w {\bf Y} \,.
\label{ES5}
\end{equation}
Here $\tau$ is a relaxation time, a new coupling parameter of the model.
In the cosmological context $\tau$ is generally the function of time, $\tau(t)$.
We assume that $ \tau(t) {=} \frac{\xi}{\sigma H(t)}$, i.e., this relaxation time can be measured in natural
cosmological scale \cite{tau1}. Our ansatz here is that the
dimensionless parameter $\xi$ is constant.

\subsubsection{Key equation of evolution of the DE pressure}

When the quantities $\rho(t)$ and $\Pi(t)$ depend on time through
the scale factor $a(t)$ only, i.e., $\rho {=} \rho(a(t))$, $\Pi =
\Pi(a(t))$, the so-called $x$-representation  is convenient, based
on the following relations
\begin{equation}
\frac{d}{dt} = x H(x)
\frac{d}{dx} \,, \quad  t-t_0 = \int_1^{\frac{a(t)}{a(t_0)}}
\frac{dx}{x H(x)} \,. \label{diffa}
\end{equation}
In these terms the balance equation (\ref{balance22}) takes the
following form (the prime denotes the derivative with respect to $x$)
$$
x \rho^{\prime}(x) + 3(\rho + \Pi) =
$$
\begin{equation}
{=} {-} \sum_{({\rm a})} E_{({\rm a})} \frac{\left[x^2
F_{({\rm a})}(x)\right]^{\prime}}{2x^4} \int_0^{\infty}\frac{q^4
dq e^{{-}\lambda_{({\rm a})} \sqrt{1{+}q^2}}}{\sqrt{1{+}q^2 F_{({\rm a})}(x)}} .
\label{RHOkey}
\end{equation}
Using (\ref{simplest0}) one can transform this equation into the equation
for the DE pressure only, yielding
\begin{equation}
\xi x^2 \Pi^{\prime \prime}(x) {+} x \Pi^{\prime}(x) \left(4 \xi {+}
\sigma \right) {+} 3 (1{+}\sigma)\Pi {+}
3 \rho_0 {=} {\cal J}(x) \,,
\label{key1}
\end{equation}
where the source ${\cal J}(x){=}{\cal J}(x,\Pi{-}\Pi(1),\Pi^{\prime})$ is defined as follows:
\begin{equation}
{\cal J}(x) \equiv {-} \sum_{({\rm a})} E_{({\rm a})} \frac{\left[x^2
F_{({\rm a})}(x)\right]^{\prime}}{2 x^4} \int_0^{\infty}\frac{q^4
dq e^{{-}\lambda_{({\rm a})} \sqrt{1{+}q^2}}}{\sqrt{1{+}q^2 F_{({\rm a})}(x)}} .
\label{key2}
\end{equation}
The quantity ${\cal J}(x)$ vanishes, when all the Archimedean parameters vanish, i.e.,
${\cal V}_{({\rm a})}=0$. Below we will address Eq. (\ref{key1}) as the {\it key equation}. It is a
differential equation of the second-order linear in the derivatives but nonlinear in the unknown
function $\Pi(x)$.
There are two important particular cases, when the sourceterm ${\cal J}(x)$ can be written in an explicit form; let us consider
them in more detail.

\subsubsection{Two explicit examples}

\noindent
{\it (i) Massless dark matter}

\noindent
When one deals with massless particles [enumerated, e.g., by the index $(0)$], the source term ${\cal J}(x)$ takes explicit
form
\begin{equation}
{\cal J}_{(0)}(x) = E_{(0)}
\frac{{\cal V}_{(0)}
 \ \Pi^{\prime}(x)}{x^3} \exp{\{{\cal V}_{(0)}
[\Pi(1){-}\Pi(x)]\}} \label{key3}
\end{equation}
with $E_{(0)}$ given by (\ref{m00}). When DM particles are massive
but effectively ultrarelativistic [$q^2 F_{({\rm a})}(x) >> 1$
and $q^2 >> 1$], the source term ${\cal J}(x)$ has the same form
(\ref{key3}).

\vspace{3mm}

\noindent
{\it (ii) Cold dark matter}

\noindent When one deals with the models using the concept of cold
dark matter, one assumes that the corresponding particles are
effectively nonrelativistic, i.e., $q^2 F_{({\rm a})} << 1$ and
$q^2 << 1$. The corresponding key equation contains the source
term
\begin{equation}
{\cal J}_{({\rm C})}(x) = \frac{3 N_{({\rm C})} T_{({\rm C})} {\cal
V}_{({\rm C})} \Pi^{\prime}(x)}{x^4} \ e^{2 {\cal V}_{({\rm C})}
[\Pi(1){-}\Pi(x)]} \,. \label{nrP00key}
\end{equation}
Ultrarelativistic and nonrelativistic models can be studied
analytically and qualitatively. When the parameters
$\lambda_{({\rm a})}$ are arbitrary, one needs numerical analysis
for the key equation.

\section{Two examples of exact solutions}

\subsection{Constant solution to the key equation}

The behavior of the function $\Pi(x)$, the solution to the Eq. (\ref{key1}), essentially depends on the initial data
$\Pi(1)$, $\Pi^{\prime}(1)$, on the values of the parameters
$\lambda_{({\rm a})}$, $\sigma$, $\rho_0$, $\xi$ and ${\cal
V}_{({\rm a})}$. When $\sigma \neq -1$, the key equation admits a
special constant exact solution
\begin{equation}
\Pi(x) = \Pi(1) = - \frac{\rho_0}{1+\sigma} = \Pi(\infty) \,.
\label{spec1}
\end{equation}
For this solution the energy density $\rho(x)$ is also constant
\begin{equation}
\rho(x) = \rho(1) = - \Pi(x) = \frac{\rho_0}{1+\sigma} = - \Pi(\infty) \,,
\label{spec11}
\end{equation}
and the parameters $\xi$, $\lambda_{({\rm a})}$ and ${\cal V}_{({\rm
a})}$ are arbitrary. For this special constant solution the
Archimedean-type force vanishes (or more precisely becomes
hidden), and the cosmological model turns into the FLRW-type model
with dark matter and nonvanishing cosmological constant
$\Lambda$. For this model we should write
\begin{equation}
\Pi(x) = - \rho(x) = - \frac{\Lambda}{8\pi G}  \,, \label{spec2}
\end{equation}
and assume that
\begin{equation}
\rho_0 = \Lambda \ \frac{(1+\sigma)}{8\pi G} \,, \label{spec3}
\end{equation}
taking into account the solution (\ref{spec1}). This constant solution also exists when
the Archimedean-type force is absent, i.e., when ${\cal V}_{({\rm a})}{=}0$ and ${\cal J}(x){=}0$.
In addition, this solution is an asymptotic limit for a family of integral curves at
$x \to \infty$.

A question arises: for what values of the parameters
$\sigma$, $\rho_0$, $\xi$ and ${\cal V}_{({\rm a})}$ is this special constant solution 
asymptotically stable? In other words, when does the deviation from the asymptotic value tend to zero, $Z(x\to \infty) \to 0$, where
\begin{equation}
\Pi(x) {=} {-} \frac{\rho_0}{1{+}\sigma} {+} Z(x) ,  \ Z(1) {=} 0 ,
\ Z^{\prime}(1) {=} \Pi^{\prime}(1) \neq 0 \,. \label{spec17}
\end{equation}
In order to answer this question let us analyze Eq.
(\ref{key1}) linearized with respect to $Z$ [see (\ref{spec17})]
at $x \to \infty$. In this case the leading-order term in the
decomposition (\ref{spec17}) satisfies the equation
\begin{equation}
\xi x^2 Z^{\prime \prime}(x) + x Z^{\prime}(x) \left(4\xi + \sigma \right) + 3 (1+\sigma) Z  = 0
\,.
\label{key17}
\end{equation}
This equation is clearly the well-known Euler equation. First,
it can be obtained as an exact solution for the case ${\cal
V}_{({\rm a})}=0$; second, this equation describes integral
curves slightly deviating from the constant solution (\ref{spec1})
at $x \to \infty$. The characteristic polynomial of the Euler
equation (\ref{key17}) has two roots
\begin{equation}
s_{1,2} = \frac{1}{2\xi} \left[- (\sigma + 3\xi) \pm \sqrt{(\sigma -
3\xi)^2 -12\xi} \right] \,, \label{roots17}
\end{equation}
which can be real or complex depending on the values of the parameters $\xi$
and $\sigma$. One can distinguish three subcases.

\subsubsection{Two different real roots [$(\sigma-3\xi)^2 > 12\xi$]}

When the discriminant in (\ref{roots17}) is positive, one obtains
\begin{equation}
Z(x) =  \frac{\Pi^{\prime}(1)}{2\Gamma} x^{-\gamma} \left(x^{\Gamma} -
x^{-\Gamma} \right)  \,, \label{aperiod2}
\end{equation}
where
\begin{equation}
\gamma \equiv \frac{\sigma + 3 \xi}{2\xi} \,, \quad \Gamma \equiv \frac{1}{2\xi} \sqrt{(\sigma
-3\xi)^2 -12\xi} \,. \label{aperiod1}
\end{equation}
The solution (\ref{aperiod2}) tends to zero asymptotically at $x \to \infty$,
when $\Gamma < \gamma$; it is possible, first, when
$\sigma > 3\xi {+}2\sqrt{3\xi}$ (for arbitrary $\xi$), second, when
${-}1 < \sigma < 3\xi {-} 2\sqrt{3\xi}$ for $\xi > \frac{1}{3}$.

\subsubsection{Double real roots [$(\sigma-3\xi)^2 = 12\xi$]}

In this case the solution for $Z(x)$ is
\begin{equation}
Z(x) =  \Pi^{\prime}(1) \ x^{-\gamma} \log{x}   \,, \label{aperiod22}
\end{equation}
where $\gamma {=} 3 {+} \sqrt{\frac{3}{\xi}}$, if $\sigma {=} 3\xi {+}
2\sqrt{3\xi}$, and $\gamma {=} 3 {-} \sqrt{\frac{3}{\xi}}$, if $\sigma
{=} 3\xi {-} 2\sqrt{3\xi}$. The first solution relates to the
asymptotically vanishing $Z(x \to \infty)$ for all $\xi$, the
parameter $\sigma$ being positive. The second solution tends to
zero asymptotically, when $\xi > \frac{1}{3}$, and the parameter
$\sigma$ satisfies the inequality $\sigma > {-}1$. Finally, when
$\xi {=} \frac{1}{3}$ and $\sigma {=} {-}1$, the solution is logarithmically
unstable.

\subsubsection{Complex roots [$(\sigma-3\xi)^2 < 12\xi$]}

For the complex roots the solution is quasiperiodic
\begin{equation}
Z(x) = \frac{\Pi^{\prime}(1)}{\beta} \ x^{-\gamma} \sin{(\beta \log{x})}  \,,
\label{period2}
\end{equation}
where
\begin{equation}
\gamma \equiv \frac{\sigma + 3
\xi}{2\xi} \,, \quad \beta \equiv \frac{1}{2\xi} \sqrt{12 \xi -
(\sigma -3\xi)^2} \,. \label{period1}
\end{equation}
This solution remains quasiperiodic and asymptotically small, when $\gamma
>0$; it is possible in the following two cases:

\noindent
(i) when $0< \xi < \frac{1}{3}$ and $-3\xi < \sigma <
3\xi + 2 \sqrt{3\xi}$;

\noindent
(ii) when $ \xi \geq \frac{1}{3} $ and $3\xi -
2 \sqrt{3\xi} \leq \sigma \leq 3\xi + 2 \sqrt{3\xi}$.

\noindent
A special case, when the roots are purely imaginary, relates to the condition $\gamma =0$ or equivalently
$\sigma = - 3\xi$. This special case is realized, when $\xi <
\frac{1}{3}$.

To conclude, we can state that the solution $\Pi(x) {=} {-}
\frac{\rho_0}{1{+}\sigma}$ is asymptotically stable, first, when $0<\xi<\frac{1}{3}$
and $\sigma > {-}3\xi$, second, when $\xi \geq \frac{1}{3}$ and $\sigma > {-}1$.
This statement is illustrated by the Fig.1.

\begin{figure}
[htmb]
\includegraphics{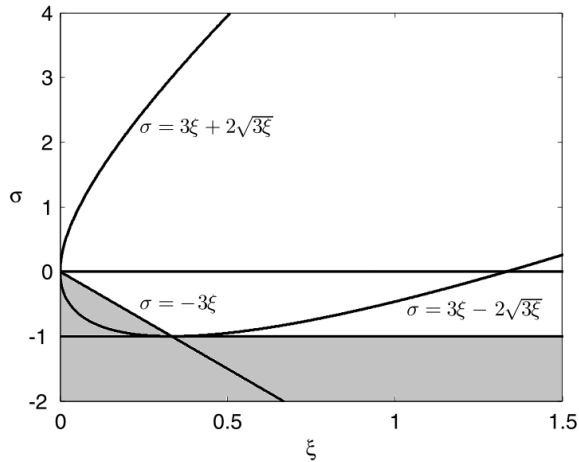}%

\caption {{\small The stability domain of the special solution
$\rho {=} {-} \Pi {=} \frac{\rho_0}{1{+}\sigma}$, presented on the plane
of the parameters $\xi$ and $\sigma$. The deviation from this
constant solution decreases according to the power law
(\ref{aperiod2}), first, in the zone situated higher than the
upper branch of the parabola, $\sigma {=} 3\xi {+} 2 \sqrt{3\xi}$ for
arbitrary $\xi$, second, in the zone between the straight line
$\sigma {=} {-}1$ and the lower branch of the parabola, $\sigma {=} 3\xi
{-} 2 \sqrt{3\xi}$ for $\xi>\frac{1}{3}$. In the zone inside the
parabola higher than the straight line $\sigma {=} {-}3\xi$ the deviation
from the constant solution behaves as a damped sinusoid
(\ref{period2}). On the branches of parabola the decreasing is
described by the product of a power-law function and logarithm
(\ref{aperiod22}).}}
\end{figure}

Since the zone of stability relates to the case $\sigma > {-}1$, the asymptotic value
of the DE pressure is negative and the dark energy is positive at $\rho_0 >0$.
In the asymptotic limit $x \to \infty$ the energy density of the
dark matter asymptotically vanishes, $E(\infty) = 0$, thus the Hubble function and the
scale factor tend to the following values
\begin{equation}
H(\infty) = \sqrt{\frac{8\pi G \rho_0}{3 (1+\sigma)}} \,, \quad
a(t) \to a(t_0) e^{H(\infty) t}\,. \label{ina3}
\end{equation}
Such a behavior corresponds to the de Sitter law with positive
acceleration parameter $-q \equiv \frac{\ddot{a}}{aH^2} = 1$,
thus, one can indicate this exact solution as a $\Lambda$ - type
solution of the model with Archimedean-type interaction between DM
and DE.

When $\sigma <{-}1$, the constant solution (\ref{spec1}) is unstable for
arbitrary $\xi$. The case $\sigma {=} {-}1$ is the special one, and we consider
this case in detail in the next section.

\subsection{Special model with $\sigma {=} {-}1$: \\ Exact solutions of the anti-Gaussian type}

For this specific model the key equation (\ref{key1}) does not
admit constant solutions if $\rho_0 \neq 0$. In order to analyze
the new situation we consider, first, massless dark matter,
then the model with cold dark matter, and discuss exact solutions
of the key equation, which now become logarithmic, providing the
scale factor to be of the anti-Gaussian type.

\subsubsection{Special model for a massless dark matter}

When $\sigma {=} {-}1$ the key equation of the second order for massless dark matter
$$
\xi x^2 \Pi^{\prime \prime}(x) {+} x \Pi^{\prime}(x) (4 \xi {-} 1){-}
$$
\begin{equation}
{-} \Pi^{\prime}(x) E_{(0)} \frac{{\cal V}_{(0)}}{x^3} \ e^{{\cal V}_{(0)}
[\Pi(1){-}\Pi(x)]} {+} 3 \rho_0  {=} 0
\label{keyanti1}
\end{equation}
can be reduced to the first-order equation
\begin{equation}
\xi z [4 + {\cal V}_{(0)} W] \frac{d W}{d z} = W \left[3\xi - 1 - E_{(0)} {\cal V}_{(0)} z
\right] + 3 \rho_0
\label{keyanti2}
\end{equation}
based on the following definitions (see, e.g., \cite{PolZa}):
\begin{equation}
y = \Pi(x) - \Pi(1) \,, \quad x \frac{dy}{dx} = W \,, \quad
z = x^{-4} e^{-{\cal V}_{(0)} y } \,.
\label{keyanti3}
\end{equation}
Eq. (\ref{keyanti2}) is the Abel equation of the second kind \cite{PolZa}.
The special solution to this equation can be obtained if we put
the coefficient $[4 + {\cal V}_{(0)} W]$ in front of the derivative
equal to zero. Two constants
\begin{equation}
W = - \frac{4}{{\cal V}_{(0)}} \,, \quad
z = \frac{1}{E_{(0)}} \left[\frac{3\xi -1}{{\cal V}_{(0)}} - \frac{3}{4} \rho_0
\right]
\label{keyanti4}
\end{equation}
satisfy the key equation, if
\begin{equation}
y(x) {=} {-} \frac{4}{{\cal V}_{(0)}} \log{x} , \quad z{=}1 , \quad
E_{(0)} {=}  \left[\frac{3\xi {-}1}{{\cal V}_{(0)}} {-} \frac{3}{4} \rho_0
\right] .
\label{keyanti5}
\end{equation}
Of course, this special solution exists when ${\cal V}_{(0)} \neq 0$, i.e.,
when the Archimedean-type force is present.

\subsubsection{Logarithmic solution for the DE pressure}

When the parameters of the model are coupled by the last relation in
(\ref{keyanti5}), we obtain an exact solution
\begin{equation}
\Pi(x) = \Pi(1) - \frac{4}{{\cal V}_{(0)}} \log{x} \,,
\label{etoy11}
\end{equation}
\begin{equation}
\rho(x) = \rho(1) + \frac{4}{{\cal V}_{(0)}} \log{x} \,,
\label{toy11}
\end{equation}
where the relation
\begin{equation}
\Pi^{\prime}(1) = - \frac{4}{{\cal V}_{(0)}}  =
\frac{1}{\xi}\left[ \rho(1) - \rho_0 + \Pi(1) \right]
\label{toy011}
\end{equation}
is used.
The sum $\Pi(x) {+} \rho(x)$ remains constant for all values of $x$
\begin{equation}
\rho(x) + \Pi(x) = \rho(1) + \Pi(1) = \rho_0 -
\frac{4\xi}{{\cal V}_{(0)}} \,. \label{toy2}
\end{equation}
The DM energy density $E(x)$ is also constant, i.e.,
\begin{equation}
E(x) = const = E_{(0)} = - \frac{3}{4} \rho_0 +
\frac{3\xi-1}{{\cal V}_{(0)}} \,, \label{toy3}
\end{equation}
due to the specific structure of the corresponding Archimedean-type force.
The Hubble function $H(x)$ can be found from Eq.
(\ref{EinREDU1}), which now takes the form
\begin{equation}
H^2(x) = \frac{8\pi G}{3} \left[\rho(1)  + E_{(0)} +
\frac{4}{{\cal V}_{(0)}} \log{x} \right] \,. \label{toy4}
\end{equation}
Since the right-hand side of this equation has to be nonnegative
for every $x>1$, we suppose that in addition to ${\cal V}_{(0)} > 0$
the following inequality holds
\begin{equation}
\rho(1)  + E_{(0)} \geq 0 \
\rightarrow \Pi(1) \leq \frac{1}{4}\rho_0 -
\frac{1+\xi}{{\cal V}_{(0)}} \,. \label{toy41}
\end{equation}
According to (\ref{diffa}) and (\ref{toy4}) the scale factor $a(t)$ can be written in the form
\begin{equation}
a(t) = a^{*} \exp\left\{\frac{8\pi G}{3 {\cal V}_{(0)}}
(t-t^{*})^2 \right\}  \,, \label{toy5}
\end{equation}
where the parameters with asterisks are defined as follows
$$
a^{*} \equiv a(t_0) \exp\left\{- \frac{{\cal
V}_{(0)}}{4}[\rho(1)+ E_{(0)}] \right\}  \,,
$$
\begin{equation}
t^{*} \equiv
t_0 - {\cal V}_{(0)} \sqrt{\frac{3}{32 \pi G}[\rho(1)+
E_{(0)}]} \,. \label{toy6}
\end{equation}
Such a scale factor describes an anti-Gaussian expansion
without initial singularity. The acceleration parameter $-q(t)$
for the anti-Gaussian expansion is positive and exceeds one:
\begin{equation}
- q(t) \equiv \frac{\ddot{a}}{a H^2} = 1 + \frac{3 {\cal
V}_{(0)}}{ 16 \pi G (t{-}t^{*})^2}  \  \geq 1 \,. \label{toy7}
\end{equation}
Moreover, it is a regular function of time (since $t>t_0>t^{*}$),
and tends to one, when $t \to \infty$. Comparing the anti-Gaussian function (\ref{toy5}) with
the corresponding de Sitter type function
\begin{equation}
a(t) = a(t^{*})  \exp\left\{\sqrt{\frac{\Lambda}{3}} (t-t^{*}) \right\} \,,
\label{deSitter}
\end{equation}
one can see that the former increases slowly at $t \simeq t^{*}$, but then grows much
more quickly.

\subsubsection{Stability analysis of the anti-Gaussian solution}

The differential equation (\ref{keyanti2}) can be evidently reduced to the
autonomous dynamic system
\begin{equation}
\frac{dz}{d \tau} =\xi z [4 + {\cal V}_{(0)} W] \,,
\label{keyanti201}
\end{equation}
\begin{equation}
\frac{d W}{d \tau} = W \left[3\xi - 1 - E_{(0)} {\cal V}_{(0)} z
\right] + 3 \rho_0 \,.
\label{keyanti202}
\end{equation}
The exact constant solution (\ref{keyanti4})
\begin{equation}
W_{*} = - \frac{4}{{\cal V}_{(0)}} \,, \quad
z^{*} = 1 \,,
\label{keyanti401}
\end{equation}
valid at the condition (\ref{keyanti5}), describes the stationary (singular) point of this
dynamic system. In order to determine the type of this singular point let us
consider the linearized system
\begin{equation}
\frac{d \zeta}{d \tau} =\xi {\cal V}_{(0)} \omega \,, \quad
\frac{d \omega}{d \tau} = \frac{3}{4} \omega \rho_0 {\cal V}_{(0)} + 4 E_{(0)} \zeta
\,,
\label{keyanti204}
\end{equation}
where
\begin{equation}
W \to W_{*} + \omega \,, \quad z \to z^{*} + \zeta \,.
\label{keyanti402}
\end{equation}
Since we consider the case, when  $\xi>0$, $E_{(0)}>0$, ${\cal
V}_{(0)}>0$, the roots of the corresponding characteristic
equation are real
\begin{equation}
s_{1,2} =  \frac{3}{8} \rho_0 {\cal V}_{(0)}
\left[ 1 \pm \sqrt{1+ \frac{256 \xi E_{(0)}}{9 {\cal V}_{(0)} \rho^2_0}} \ \right]\,.
\label{keyanti403}
\end{equation}
The product of the roots $s_1 \cdot s_2 {=} {-} 4 \xi E_{(0)} {\cal V}_{(0)} < 0$ is negative, thus,
the singular point is the saddle one, and the anti-Gaussian solution is
unstable.

\subsubsection{Cold dark matter}

The key equation (\ref{key1}) with the nonrelativistic source
(\ref{nrP00key}) can also be reduced to the first-order equation
\begin{equation}
\xi z [5 {+} 2 {\cal V}_{({C})} W] \frac{d W}{d z} {=} W \left[3\xi {-} 1 {-} 3 N_{({C})} T_{({C})} {\cal V}_{({C})} z
\right] {+} 3 \rho_0
\label{keyanti21}
\end{equation}
based on the following definitions:
\begin{equation}
y {=} \Pi(x) {-} \Pi(1) \,, \quad x \frac{dy}{dx} {=} W \,, \quad
z {=} x^{{-}5} e^{{-}2{\cal V}_{({C})} y } \,.
\label{keyanti31}
\end{equation}
The corresponding exact special solution is
\begin{equation}
\Pi(x) = \Pi(1) - \frac{5}{2{\cal V}_{({C})}} \log{x} \,,
\label{NRtoy1}
\end{equation}
when
\begin{equation}
\frac{6}{5} \rho_0 + 3 N_{({C})} T_{({C})} = \frac{1}{{\cal V}_{({C})}} (3\xi-1) \,.
\label{NRtoy03}
\end{equation}
The corresponding value of the DE energy density is
\begin{equation}
\rho(x) = \rho(1) + \frac{5}{2{\cal V}_{({C})}} \log{x} \,,
\label{NRtoy11}
\end{equation}
so, that the sum of $\Pi(x)$ and $\rho(x)$ again remains constant
for all values of $x$
\begin{equation}
\rho(x) + \Pi(x) = \rho(1) + \Pi(1) = \rho_0 - \frac{5\xi}{2{\cal
V}_{({C})}} \,. \label{NRtoy2}
\end{equation}
The DM energy-density function $E(x)$ is now a decreasing function of $x$ (not a constant contrary to
the massless case)
\begin{equation}
E(x) = \frac{m_{({C})} N_{({C})}}{x^3} \,, \label{NRtoy3}
\end{equation}
and the Hubble function $H(x)$ can be now found from the equation
\begin{equation}
H^2(x) = \frac{8\pi G}{3} \left[\rho(1)  +  \frac{5}{2{\cal V}_{({C})}}
\log{x} + \frac{m_{({C})} N_{({C})}}{x^3}\right] \,. \label{NRtoy4}
\end{equation}
In the asymptotic regime $x \to \infty$ the logarithmic term in
(\ref{NRtoy4}) dominates, thus, we obtain again the anti-Gaussian
solution
\begin{equation}
a(t) = a^{**} \exp\left\{\frac{20 \pi G}{3{\cal V}_{({C})}} (t-t^{**})^2
\right\}  \,, \label{NRtoy5}
\end{equation}
where the parameters with double asterisks are defined as follows
$$
a^{**} \equiv a(t_0) \exp\left\{- \frac{\rho(1){\cal V}_{({C})}}{5}
\right\} \,,
$$
\begin{equation}
 t^{**} \equiv t_0 - \frac{{\cal V}_{({C})}}{10}
\sqrt{\frac{3\rho(1)}{ 2\pi G}} \,. \label{NRtoy6}
\end{equation}
Again it will be an accelerated expansion of the Universe with ${-}q>1$, and this solution is also
unstable.

\subsubsection{Model with DE domination}

In order to complete the analysis of the model with $\sigma {=} {-}1$, let us consider
the case, when the dark matter is absent ($E_{({\rm a})}{=}0$).
In fact such a model can be considered as the approximate one, since at $x \to
\infty$ the DM contribution to the total energy decreases as $x^{{-}3}$.
The key equation
for the DE pressure reduces now to the inhomogeneous Euler equation
\begin{equation}
\xi x^2 \Pi^{\prime \prime}(x) + x \Pi^{\prime}(x) \left(4\xi - 1
\right) + 3 \rho_0 = 0 \,. \label{keyinfty}
\end{equation}
The characteristic equation for the corresponding homogeneous Euler equation
\begin{equation} s (s - \nu) = 0 \,,  \label{Euler1}
\end{equation}
with $\nu \equiv \frac{1{-}3\xi}{\xi}$, gives double roots $s{=}0$, when $\xi {=} 1/3$,
thus, let us consider two different cases.

\noindent
{\it (i) Special case: $\sigma = -1$, $\xi \neq \frac{1}{3}$.}

\noindent
Exact solution to (\ref{keyinfty}) contains in this case a sum of
logarithmic and power-law terms
$$
\Pi(x) {=} \Pi(1) {+} \frac{3\rho_0 \log{x}}{1{-}3\xi}
{+} \frac{(x^{\nu}{-}1)}{1{-}3\xi} \left[\rho(1){+}\Pi(1) {+} \frac{\rho_0}{3\xi{-}1}\right],
$$
\begin{equation}
\rho(x) {=} \rho(1) {-} 3\xi \left[\Pi(x){-}\Pi(1) \right] {-} 3\rho_0 \log{x} \,. \label{Sc2}
\end{equation}
Asymptotic behavior of the
DE pressure and DE energy density at $3\xi > 1$ for
arbitrary initial data are the following
$$
\Pi(x \to \infty) \to  - \frac{3\rho_0}{3\xi-1} \log{x} \,,
$$
\begin{equation}
\rho(x \to \infty) \to  \frac{3\rho_0}{3\xi -1} \log{x} \,.
\label{Sc3}
\end{equation}
In the asymptotic regime we deal again with the anti-Gaussian
expansion
\begin{equation}
a(t \to \infty) \propto  \exp\left\{\frac{2\pi G \rho_0}{(3\xi-1)}
t^2 \right\} \,,
\label{Sc4}
\end{equation}
when $\xi > \frac{1}{3}$. We would like to mention that when the
initial data $\Pi(1)$, $\rho(1)$ and $\Pi^{\prime}(1)$ take special values
\begin{equation}
\Pi(1) = {-} \frac{\rho_0}{3\xi{-}1} \,, \quad \rho(1) = 0 \,, \quad
\Pi^{\prime}(1) = {-} \frac{3\rho_0}{3\xi{-}1} \,, \label{Gauss1}
\end{equation}
the power-law terms disappear from the exact formulas
\begin{equation}
\Pi(x) = \frac{\rho_0}{1{-}3\xi} (1{+}3\log{x}) \,, \quad \rho(x) =
 \frac{3\rho_0}{3\xi{-}1} \log{x} \,, \label{Gauss2}
\end{equation}
and the anti-Gaussian solution
\begin{equation}
a(t) = a(t_0) \exp\left\{\frac{2\pi G \rho_0}{(3\xi-1)} (t-t_0)^2
\right\} \,, \label{Gauss3}
\end{equation}
become not only asymptotic, but the exact one.
When $\nu > 0$, then $\rho(x) \propto x^{\nu}$,  $H \propto
x^{\frac{1}{2}\nu}$ and $a(t \to \infty) \to 0$. Thus, the Universe does not
expand, when $\xi < \frac{1}{3}$.

\noindent
{\it (ii) Special case: $\sigma {=} {-}1$, $\xi {=} \frac{1}{3}$.}

The root $s{=}0$ of characteristic Eq.(\ref{Euler1}) is now double,
so, the exact solution contains the logarithmic terms in square,
\begin{equation}
\Pi(x) {=} \Pi(1) {+} 3 \left[\rho(1) {+} \Pi(1) {-} \rho_0 \right] \log{x} {-} \frac{9}{2}\rho_0 \log^2{x} \,, \label{Sc31}
\end{equation}
\begin{equation}
\rho(x) = \rho(1) - 3 \left[\rho(1) + \Pi(1) \right] \log{x} + \frac{9}{2}\rho_0 \log^2{x} \,, \label{Sc41}
\end{equation}
the sum of these functions being linear in logarithm,
\begin{equation}
\Pi(x) + \rho(x) =  \Pi(1) + \rho(1) - 3 \rho_0 \log{x} \,.
\label{Sc51}
\end{equation}
The scale factor evolves now superexponentially. For instance, when the initial data satisfy the condition
$\rho(1){+}\Pi(1){=}0$, the scale factor has the form
\begin{equation}
\frac{a(t)}{a(t_0)} {=} \exp\left\{ \sqrt{\frac{2\rho(1)}{9\rho_0}} \sinh{\left[\sqrt{12\pi G \rho_0} (t{-}t_0) \right]} \right\}. \label{Sc61}
\end{equation}
Near the starting point $t_0$ the function $a(t)$ behaves according to the de Sitter law
\begin{equation}
a(t) \simeq a(t_0) e^{H(t_0)(t-t_0)} \,, \quad H(t_0) = \sqrt{\frac{8\pi G \rho(1)}{3}} \,,
\label{Sc71}
\end{equation}
while asymptotically at $t \to \infty$ the scale factor grows as
\begin{equation}
a(t) \to a(t_0) \exp\left\{ \sqrt{\frac{2\rho(1)}{9\rho_0}} \ e^{\sqrt{12\pi G \rho_0} \ (t{-}t_0) } \right\} . \label{Sc65}
\end{equation}
For the Universe expansion described by the law (\ref{Sc61}) the Hubble function is monotonic
\begin{equation}
H(t) = \sqrt{\frac{8\pi G \rho(1)}{3}} \cosh{\left[\sqrt{12\pi G \rho_0} \ (t-t_0) \right]} \label{Sc67}
\end{equation}
and increases infinitely. The acceleration parameter
\begin{equation}
-q(t) = 1 {+} \sqrt{\frac{9\rho_0}{2\rho(1)}} \ \frac{\sinh{\left[\sqrt{12\pi G \rho_0} \ (t{-}t_0) \right]}}{\cosh^2{\left[\sqrt{12\pi G \rho_0} \ (t{-}t_0) \right]}} \label{Sc69}
\end{equation}
starts with ${-}q(t_0){=}1$, reaches the maximum ${-}q_{({\rm max})}{=}1 {+} \sqrt{\frac{9\rho_0}{8\rho(1)}}$ at $t{=}t_0 {+} \frac{\log{(1+\sqrt2)}}{\sqrt{12\pi G \rho_0}}$, and tends
asymptotically to ${-}q(\infty){=}1$.

\subsubsection{Remark on the coupling of the baryon matter to DE}

The baryon matter can be naturally included into the scheme of Archimedean-type coupling: for this purpose one can add to the sum the terms, which correspond 
to the standard particles. If one assumes that the baryon matter is not affected by the Archimedean-type coupling, one can put the corresponding coupling constants
${\cal V}_{({\rm a})}$ equal to zero; nevertheless, the contribution of the baryon matter to the total stress-energy tensor now will be taken into account in the modified formula (\ref{TEI}).

\section{Discussion}

The study of the cosmological model, into which the
Archimedean-type interaction between dark energy and dark matter
is introduced, shows that the roles of DM and DE in the energy
balance of the Universe can be revised. According to the obtained
formula (\ref{E(x)}), the contribution of the DM particles,
$E_{({\rm a})}(a(t))$, into the total energy density  depends on
the state of DE pressure $\Pi(a(t))$ through the functions
$F_{({\rm a})}(x)$ (see (\ref{Fdefin})). In the models without
Archimedean-type force the energy of the DM particle decreases
effectively because of the factor $a^{-2}(t)$; in other words, all
the particles, both nonrelativistic and ultrarelativistic at the
initial moment $t_0$, inevitably become (effectively)
nonrelativistic in the process of Universe expansion. When the
Archimedean-type force acts on the DM particles, the particle
energy (\ref{energy}) becomes much more complicated function of
cosmological time due to the exponential dependence on the DE
pressure. This means, in particular, that, nonrelativistic
particles can become (effectively) ultrarelativistic due to the
Archimedean-type force action, thus the corresponding contribution
of cold dark matter into the total energy can be reestimated
taking into account the sign, the value of the DE pressure at this
moment, as well as the rate of its variation with time. In
contrast, the ultrarelativistic DM particles  can
become (effectively) nonrelativistic, when the corresponding
exponential factor in (\ref{energy}) is small. Now the
contribution of cold dark matter into the total energy is
estimated to be about $23 \%$. The question arises: does this
estimate include a total rest energy of the massive DM particles
only, or the energy of the Archimedean-type interaction as well? We
need such a clarification, for instance, in order to estimate
the density numbers of the DM particles of different sorts; in
particular, the information about the number density of DM axions
in the terrestrial laboratories is very important for planning
experiments with axion electrodynamics (see, e.g., \cite{WTNi}).

\vspace{3mm}

The cosmological model with Archimedean-type force
describes a self-regulating Universe. This means that in the process
of expansion of the Universe the total (conserved as a whole)
energy can be redistributed between dark energy and dark matter
constituents according to the challenge of the corresponding
epoch. The energy pendulum stimulated by the Archimedean-type force
can work by the following scheme: let us imagine that at some
moment the DE pressure $\Pi$ is negative and is varying rather
quickly; then according to the formulas (\ref{key2}),
(\ref{key3}), (\ref{nrP00key}) the DM particle reaction, provoked
by the Archimedean-type force, will be strong. The
corresponding intensive source appears in the right-hand side of
the key equation (\ref{key1}), thus decreasing the rate of $\Pi$
evolution. From the theoretical point of view, it is not yet
clear, first, for which set of guiding parameters such a specific
regime does exist; second, when such a regime can be characterized as
(quasi)oscillations; and third, how the number of epochs of the
Universe expansion does depend on the set of guiding parameters of the
model. We started to study these questions qualitatively and
numerically in the second part of our work and presented examples
of quasi-periodic, multi-inflationary and multistage evolutionary
schemes in the framework of the model based on the
Archimedean-type interaction between dark energy and dark matter.

\vspace{3mm}

In the first part of the work we focused on exact solutions of this new model. The first exact
solution is the constant one, $\Pi(x) {=} {-} \rho(x) {=} {-}\frac{\rho_0}{1{+}\sigma}$,
and relates to the case $\sigma \neq {-}1$. Since the DE pressure for this exact
solution is constant, the Archimedean-type force becomes hidden, and we obtain the
standard cosmology with $\Lambda$ term. This solution is asymptotically
stable, when the guiding parameters of the model, $\xi$ and $\sigma$,
satisfy the inequalities $0<\xi<\frac{1}{3}$ and $\sigma > {-}3\xi$, or
$\xi \geq \frac{1}{3}$ and $\sigma > {-}1$. When $\sigma < {-}1$, the solution
is asymptotically unstable for arbitrary parameter $\xi$.

\vspace{3mm}

Exact solutions of the second class, with $\sigma=-1$, are much more sophisticated, since the
DE pressure is described by the logarithmic function of the ratio
$a(t)/a(t_0)$. The corresponding scale factor $a(t)$ is given by the
anti-Gaussian function (\ref{toy5}), it has no singular points in the early Universe,
and describes late-time accelerated expansion with the acceleration parameter ${-}q>1$; this acceleration
parameter is bigger than for the de Sitter model. Exact solutions of the anti-Gaussian type happen to be
typical for different physical situations: for massless and massive nonrelativistic DM, for the
case with DE domination, etc. This solution is unstable.

\vspace{3mm}

There are two specific values of the guiding parameters of the model:
$\sigma {=} {-}1$ and $\xi {=} \frac{1}{3}$. The first one, $\sigma {=} {-}1$, can be
associated with the so-called phantom divider $w(t) {=} {-}1 {=}
\frac{1}{\sigma}$. The second value, $\xi {=} \frac{1}{3}$, can be denoted as
a resonance value of the relaxation time parameter. Indeed, according to the
formula (\ref{simplest0}) the function $\frac{\xi}{H(t)}$ plays a role of
the relaxation time for the DE pressure $\Pi(t)$, say, $\tau_{\Pi}$.
When $\xi {=} \frac{1}{3}$, one obtains that $\tau_{\Pi} {=} \frac{1}{3H(t)} {=}
\frac{1}{\Theta(t)}$, where $\Theta(t) {=} 3 H(t) {=} \nabla_k U^k$ is the expansion
parameter. Thus, the characteristic time of expansion $\frac{1}{\Theta(t)}$ coincides with
the relaxation time parameter for the DE pressure, introducing some
specific resonance condition. When $\sigma {=} {-}1$ and $\xi {=} \frac{1}{3}$ simultaneously, there exists
superexponential solution of the model, described by the scale factor (\ref{Sc61}).

\vspace{3mm}

A physical origin of the Archimedean-type force is not yet clear; nevertheless,
this force seems to be very interesting from the viewpoints of a new model of interaction and a new
scheme of redistribution of the cosmic energy between the interacting DE and DM constituents.
The presented model is self-consistent, simple from the point of view of analysis and very promising.
We keep in mind the story of the Chaplygin gas model \cite{Chap}: starting from a classical analogy a new evolutionary
model has been elaborated and applied to cosmology, although physical meaning of the Chaplygin scheme of
interaction is under discussion till now.

\vspace{5mm}
\begin{acknowledgments}
The authors are grateful to Professor W. Zimdahl for fruitful discussions, comments, and advice.
This work was partially supported by the Russian Foundation for Basic Research
(Grants No. 08-02-00325-a and 09-05-99015) and by Federal Targeted Programme "Scientific and Scientific-Pedagogical Personnel of the Innovative Russia"
(Grants No 16.740.11.0185 and  14.740.11.0407).
\end{acknowledgments}


\begin{thebibliography}{50}


\bibitem{DE1} E.J. Copeland, M. Sami and S. Tsujikawa, Int. J. Mod. Phys.
D {\bf 15}, 1753 (2006).

\bibitem{DE2} J. Frieman, M. Turner and D. Huterer, Ann. Rev. Astron. Astrophys. {\bf 46}, 385 (2008).

\bibitem{DE3} T. Padmanabhan, Gen. Relat. Grav. {\bf 40}, 529 (2007).

\bibitem{DM1} A. Del Popolo,  Astronomy Reports. {\bf 51}, 169 (2007).

\bibitem{DM2} G. Lazarides, Lect. Notes Phys. {\bf 720}, 3 (2007).

\bibitem{DM3} J. Silk, Lect. Notes Phys. {\bf 720}, 101 (2007).

\bibitem{A1} S.J. Perlmutter et. al., Nature. {\bf 391}, 51 (1998).

\bibitem{A2} A.G. Riess et al.,  Astron.J. {\bf 116}, 1009 (1998).

\bibitem{Sp1} E. Battaner and E. Florido, Fund. Cosmic Phys. {\bf 21}, 1 (2000).

\bibitem{Sp2} Y. Sofue and V. Rubin,  Ann. Rev. Astron. Astrophys. {\bf 39}, 137 (2001).

\bibitem{DF1} J. Ren and Xin-He Meng, Int. J. Mod. Phys. D {\bf 16}, 1341 (2007).

\bibitem{Oreview} S. Nojiri and S.D. Odintsov, Unified cosmic history in modified gravity:
from F(R) theory to Lorentz non-invariant models, arXiv:1011.0544.

\bibitem{DF2}  S. Nojiri and S.D. Odintsov, Phys. Lett. B {\bf 649}, 440 (2007).

\bibitem{DF3} I. Brevik, E. Elizalde, O. Gorbunova and A. V. Timoshkin,
Eur. Phys. J. C {\bf 52}, 223 (2007).

\bibitem{DF4}  A. Arbey, Open Astronomy Journal. {\bf 1}, 27 (2008).

\bibitem{DF5} W.S. Hipolito-Ricaldi, H.E.S. Velten and W. Zimdahl,
JCAP. {\bf 0906}, 016 (2009).

\bibitem{M1} W. Zimdahl, D. Pav\'on and L.P. Chimento, Phys. Lett. B {\bf 521}, 133 (2001).

\bibitem{M2} W. Zimdahl and D. Pav\'on,  Gen. Relat. Grav. {\bf 33}, 791 (2001).

\bibitem{M3} L.P. Chimento and  D. Pav\'on, Phys. Rev. D {\bf 73}, 063511 (2006).

\bibitem{M4} N. Cruz, S. Lepe and F. Pena, Phys. Lett. B {\bf 663}, 338 (2008).

\bibitem{M5} J. Valiviita, E. Majerotto and R. Maartens, JCAP. {\bf 0807}, 020 (2008).

\bibitem{M6} O. Bertolami, F. Gil Pedro, M. Le Delliou, Phys.Lett.B {\bf 654}, 165 (2007).

\bibitem{BZ1} W. Zimdahl and A.B. Balakin, Phys. Rev. D {\bf 58}, 063503 (1998).

\bibitem{BZ2} W. Zimdahl and A.B. Balakin, Class. Quantum Grav. {\bf 15}, 3259 (1998).

\bibitem{BZ3} W. Zimdahl, D.J. Schwarz, A.B. Balakin and D. Pav\'on,
Phys. Rev. D {\bf 64}, 063501 (2001).

\bibitem{BZ4} A.B. Balakin, D. Pav\'on, D.J. Schwarz and W. Zimdahl. New J. Phys. {\bf 5}, 85 (2003).

\bibitem{BZ5} A.B. Balakin, Gen. Relat. Grav. {\bf 36}, 1513 (2004).

\bibitem{BZ6} A. Balakin, R.A. Sussman and W. Zimdahl,  Phys. Rev. D {\bf 70}, 064027 (2004).

\bibitem{Penrose} V.G. Gurzadyan and R. Penrose, Concentric circles in WMAP data may provide evidence of violent pre-Big-Bang activity,
arXiv: 1011.3706.

\bibitem{kin1} J.M. Stewart, {\it Non-equilibrium Relativistic Kinetic Theory} (Springer, New York,
1971).

\bibitem{kin2} S.R. de Groot, W.A. van Leeuwen and Ch. G. van Weert, {\it Relativistic Kinetic
Theory} (North Holland, Amsterdam, 1980).

\bibitem{EOS1} S. Nojiri and S.D. Odintsov, Phys. Lett. B {\bf 639}, 144 (2006).

\bibitem{EOS2} V.F. Cardone, C. Tortora, A. Troisi and S. Capozziello, Phys.Rev. D {\bf 73}, 043508 (2006).

\bibitem{EOS3} S. Nojiri and S.D. Odintsov, Phys. Rev. D {\bf 72}, 023003 (2005).

\bibitem{EOS4} I. Brevik, O.G. Gorbunova and A.V. Timoshkin,  Eur. Phys. J. C {\bf 51}, 179 (2007).

\bibitem{EOS5} W. Chakraborty and  U. Debnath, Phys. Lett. B {\bf 661}, 1
(2008).

\bibitem{wt1} H. Stefancic, Phys.Rev. D {\bf 71}, 124036 (2005).

\bibitem{bag} E.V. Shuryak, Phys. Rep. {\bf 61}, 71 (1980).

\bibitem{REO1} D. Jou, J. Casas-V\'azquez and G. Lebon, {\it Extended
Irreversible Thermodynamics} (Springer, Berlin, 1996).

\bibitem{tau1} W. Zimdahl, Phys. Rev. D {\bf 61}, 083511 (2000).

\bibitem{PolZa}
A.D. Polyanin and V.F. Zaitsev, {\it Handbook of exact solutions for
ordinary differential equations} (Chapman-Hall, Boca Raton, 2000).

\bibitem{WTNi} Wei-Tou Ni, Prog. Theor. Phys. Suppl. {\bf 172}, 49 (2008).

\bibitem{Chap} A.Y. Kamenshchik, U. Moschella and V. Pasquier, Phys. Lett. B {\bf 511}, 265 (2001).

\end{thebibliography}
\end{document}